\documentclass[12pt]{iopart}
\usepackage{graphicx}
 \usepackage{amssymb}

\begin{document}

\title[  Cosmology   with   dark energy  decaying ]{ Cosmology with
dark energy decaying
 through its chemical-potential contribution }

\author{J Besprosvany}

\address{Instituto de F\'{\i}sica, Universidad Nacional Aut\'onoma de M\'exico,
Apartado Postal 20-364, M\'exico 01000, Distrito Federal, M\'exico}
\ead{bespro@fisica.unam.mx}
\begin{abstract}

 The consideration of dark energy's  quanta, required also
by thermodynamics, introduces its chemical potential into the
cosmological equations. Isolating its main contribution, we obtain
solutions with  dark energy decaying to matter or radiation. When
dominant, their energy densities  tend asymptotically to a constant
ratio, explaining today's dark energy-dark matter coincidence, and
in agreement with supernova redshift data.

\end{abstract}

 \vspace{2pc}

 \vspace{2pc}
 \noindent{\it Reference}:    J. Phys.
A: Math. Theor. 40 7099-7104 (2007) doi: 10.1088/1751-8113/40/25/S68


\section{Introduction}
Dark energy  is a    component of the universe whose
 negative pressure,  characteristic of  the quantum vacuum,
 accelerates
 its
 expansion. Evidence   for its existence
 has recently accumulated from
 independent sources as  the  supernova redshift far-distance
relation $\cite{Perlmutter},\cite{Garnavich}$,  structure
 formation$\cite{structure}$,
the microwave background radiation$\cite{microwave}$, and
lensing$\cite{lensing}$.

 The cosmological constant  $\Lambda$, dark-energy's original conception, was added by Einstein
in the application of general relativity to cosmology in 1917 in
order to describe a static universe$\cite{Einstein}$, building on a
1890s proposal  by Neumann and Seeliger, who introduced it in a
Newtonian framework for the same reasons. Its contribution in the
Einstein equations
\begin{eqnarray} \label  {equationRelativity}
R_{\mu\nu}-\frac{1}{2}g_{\mu\nu}R-\Lambda g_{\mu\nu}=8 \pi
T_{\mu\nu}
\end{eqnarray}
equilibrates gravity's attraction in a matter universe; here
$R_{\mu\nu}$ is the Ricci tensor, $g_{\mu\nu}$ the metric tensor,
which describe the geometry, and $T_{\mu\nu}$
 is the energy-momentum tensor; we use units with  the Newton, Planck, Boltzmann, and   light-speed constants
$G=\hbar=k_B=c=1,$
 except when given explicitly, as needed.

 Zeld'ovich sought to connect it to the quantum vacuum$\cite{Zel'dovich}$.
 This requires  its  reinterpretation  as a $T_{\mu\nu}$ component
 in Eq. \ref{equationRelativity}. The vacuum energy
density   of   particle fields with mass $m\ll
M_P=\frac{1}{\sqrt{G}}$ is obtained
 by summing over its modes $\bf k$:
\begin{eqnarray} \label {PlanckSpec}
   \rho_{\Lambda P}  =\frac{1}{(2 \pi)^3}\int^{M_P}d^3k
\sqrt{k^2+m^2}\simeq 3 \times 10^{114} \ {\rm \frac{GeV}{cm^3}};
\end{eqnarray}
 the
natural cutoff is the Planck-mass scale $M_P$,  the only possible
mass  conformed of $G$, $\hbar$, and $c$,
while in today's universe $\rho_{\Lambda0}\simeq 4\times  10^{-6}\
{\rm GeV/cm^3}.$
$\rho_{\Lambda0}$ represents $\Omega_{\Lambda 0}=\rho_{\Lambda
0}/\rho_{c0}\simeq .73$ of its
  critical energy density $\rho_{c0}$ today$\cite{ParticleData}$,
and in a flat universe\cite{Peebles}  $\sum \Omega_i=1$. The  rest
corresponds mainly to matter, dark and baryonic,  the latter
conforming $\Omega_{b0}\simeq
 .044$
only$\cite{ParticleData}$.  Dark energy's  origin, its smallness by
122 orders of magnitude  with respect to the vacuum's natural Planck
scale,  and the coincidence of its present energy-density scale with
the
 universe's  remain puzzling;  dynamic behavior points to a possible
 explanation.

The energy components are generally perfect fluids, described by
their energy tensor $T_{\ \  \nu}^{\mu (i)} =(\rho_i, p_i,p_i,p_i)$
(at rest), with $T_{\mu\nu}=\sum_i T_{\mu\nu}^{(i)}$.  Radiation and
matter are characterized by an equation of state
\begin{eqnarray} \label  {equationofstate} p_i=w_i \rho_i,
\end{eqnarray}
where  $w_r=1/3$  for radiation (and for  relativistic Fermi or Bose
gases,) and $w_m=0$ for non-relativistic matter.
 Under the
isotropic Robertson-Walker metric
 $ds^2=dt^2-R^2(t)(dx^2+dy^2+dz^2)$,
 Eq. \ref{equationRelativity} implies the  Friedmann equation
 \begin{eqnarray}\label  {Friedmann}
H^2 = \frac{8       \pi  }{3}\rho_c= \frac{8       \pi
}{3}(\rho_\Lambda+\rho_r+\rho_m),\end{eqnarray} where $x,y,z$ are
commoving Cartesian coordinates,  $R$ is the scale factor, depending
on time $t$, as do  $\rho_i$,
   and $H =   \dot R/R$,   the  Hubble parameter
(a dot denotes time derivative.)

 The
energy-conservation equation within    an expanding volume $V\sim
R^3$ \begin{eqnarray}\label  {energyeqCons} \sum_i d(\rho_i
V)=-\sum_i  p_i dV
\end{eqnarray}  is
implied by the contraction of Eq. \ref{equationRelativity}. When
decoupled,  each contribution also satisfies
\begin{eqnarray}\label  {energyeq} d(\rho_i V)=-p_i dV.
   \end{eqnarray}
Eq.  \ref{energyeq} can also be interpreted as  a particular case of
the first law of thermodynamics
\begin{eqnarray}\label  {EnergyConserv}
 d (\rho V)=-p dV+\mu dN+T dS,  \end{eqnarray}
with additional contributions from  the entropy $S$, and the
particle number $N$,
 where $T$ is the  temperature  and $\mu$ the
chemical potential.  When non-interactive, radiation has $\mu =0$,
baryonic matter is conserved, $dN=0$, and for both $dS=0$. These
conditions may not be true
 for dark energy or dark matter. In this   paper,  we show that
the consideration of  dark-energy's quanta modifies  the
cosmological equations through the $\mu dN$ term in Eq. \ref
{EnergyConserv},  with the implication that dark energy decays to
another component.  Thus,
 the derived asymptotic energy-density constant  ratio of the dominant components
 reproduces  the coincidence of dark energy and dark matter today.
 The entropy term $ T dS $ in Eq. \ref{EnergyConserv}  will be
 neglected, as dark energy is associated  to low-energy states.
We first classify the chemical potential associated to the pressure
in
 Eq. \ref{equationofstate} (Section 2.)
Relating it to a decay width, we consider its main contribution to
the cosmological equations, which are exactly solved for two
components, and we then apply the model to the supernova data
(Section 3.) We finally draw conclusions (Section 4.)



\section{Dark-energy's equation of state}

The
 form of  Eq. \ref{PlanckSpec} implies  $\Lambda$ generates
 a pressure $p_\Lambda=-\rho_\Lambda$, so $w_\Lambda=-1$ for the vacuum
 energy.   The parametric extension to arbitrary
negative values $w_\Lambda$,  following Eq. \ref{equationofstate},
with similar properties$\cite{Steinhardt},\cite{TurnerSmooth}$,
suits the lack of precise knowledge about it.
 Whatever is  its nature,
 and with a name not bound to its constancy,   dark energy    should contain
quanta$\cite{Zel'dovich}$, as any other form of energy in the
universe,  and so, the energy dependence on its number $N$ should be
accounted for. Within  the relation
\begin{eqnarray}\label {energyVdep} E=c V^{-w},\end{eqnarray}
 consistent with Eq. \ref{equationofstate},
$c$ a constant,   if the energy dependence  remains extensive,
another such quantity is required. Using $N$  for such a variable,
\begin{eqnarray}\label{energyVNdep}
E=c^\prime N{\left(\frac{V}{N}\right )}^{-w}
 \end{eqnarray}
  introduces an $N$-dependence, with $c^\prime$ an  (intensive)
constant, except  in the $w=-1$ case, in agreement with the view
that no quanta are associated to the vacuum.

  Eq. \ref{energyVNdep}, also consistent
with Eq. \ref{equationofstate}, implies the contribution
\begin{eqnarray}\label {chemicalPotwrho}
n\mu=(1+w)\rho,
 \end{eqnarray} where
  $n=N/V$ is  the particle density.

We concentrate on dark energy satisfying Eq. \ref{equationofstate}.
 Using the thermodynamic relation
\begin{eqnarray}
\label {thermo}
  s=\frac{1}{T}(\rho+p-n \mu ), \end{eqnarray}
  with $s=S/V$  the    entropy density, we identify two limiting cases: (1)
in the zero-entropy regime ($s=0$),
\begin{eqnarray}
\label{chemipotezerotemp} \rho_{\Lambda w}=c_{w }  n^{1+w},
\end{eqnarray}
with $c_{ w }$ a constant, and
  $n\mu_{\Lambda w}=(1+w)\rho$,
 as for Eq. \ref{chemicalPotwrho};
(2) the radiation-like assumption, $\mu_{r w}=0$,
 leads  to
\begin{eqnarray}\label {radiationlike}
 s_{r w}=c_{r
w}\rho^{\frac{1}{1+w}}
\end{eqnarray}
($c_{r w}$      a constant.)

Case (1) with  Eq. \ref{chemicalPotwrho}, induced from Eq.
\ref{equationofstate}, or Case (2) with $\mu_{r w}=0$ represent
special conditions; similarly to  Eq. \ref{equationofstate},   the
most general linear $\rho$-dependence for the chemical potential
requires the new parameter $\chi$ in
\begin{eqnarray}\label {chemipoten}
 n\mu_{w\chi}=(1+w+\chi) \rho. \end{eqnarray}
 Eqs.  \ref{equationofstate}, \ref{thermo}, and \ref{chemipoten}
then generally
 lead  to
$s_{w\chi}= n(\frac{\rho}{c_w n^{1+w}})^{-\frac{1}{\chi}}.$ From the
resulting
  temperature $T_{w \chi}=-\frac{\chi \rho} { n} (\frac{\rho}{c_w
n^{ 1+w}})^{{\frac{1}{\chi}}},$   it follows   that   $\chi\neq 0$
signals a non-zero $T_{w \chi}$.   In fact,   $s_{w\chi}$ contains
the
  $s=0$ limit, as  Eq. \ref{chemipotezerotemp}  is
approached with $\rho\sim\rho_{\Lambda w}$ for $\chi\rightarrow 0$,
  and for
 the   $\mu_{r w}=0$  case in  Eq. \ref{radiationlike},   $s_{w\chi}=s_{r
w}$ for $\chi=-w-1,$ and $c_{r w}=c_w^{1/\chi}$.
 The knowledge of $w_\Lambda$, and these   limits suggest
$\chi$ is also O(1).

The modification of Eq. \ref{energyeq} by the  chemical-potential
contribution is analyzed next.

\section{Cosmological equations with dark energy's chemical potential}
The chemical potential can be written as
\begin{eqnarray}\label {chemipotenExpand}
 \mu_\Lambda dN=
 \mu_\Lambda (n_\Lambda dV+V dn_\Lambda);
\end{eqnarray}
  changes in particle numbers through
 decay are associated to partial widths $\Gamma$, and, ultimately, to interactions.
 In the universe's evolution  in   $dt$, we distinguish the two
 contributions: (1)
 $ N \Gamma_1 dt =n_\Lambda\mu_\Lambda   dV=(1+w_\Lambda+\chi)\rho_\Lambda dV$
 is associated with   decay due to its expansion
\begin{eqnarray}\label {Gamdos}n_\Lambda \Gamma_1=3(1+w_\Lambda+\chi) H
\rho_\Lambda \sim  \rho_\Lambda^{3/2}, \end{eqnarray} given $H\sim
\rho_\Lambda^{1/2}; $ (2)
 $ N \Gamma_2 dt =\mu_\Lambda  V dn_\Lambda$ contains terms
that are not of this form; it could account for any other
out-of-equilibrium conceivable decay process linked to interactions.
For the gravitational interaction, and $T_{w\chi }=0$, $\Gamma_2\sim
\sigma n_\Lambda v \sim (1/M_P^4) n_\Lambda\rho_\Lambda^{1/2}$,
where for the cross section $\sigma \sim(1/M_P^4)
\rho^{1/2}_\Lambda$, given a tree-level gravitational interaction,
and the dimensionally fit power of the only relevant variable
$\rho_\Lambda$; the velocity $v\sim c=1$,  so
 $ n_\Lambda \Gamma_2 \sim \rho_\Lambda^{\frac{2}{w_\Lambda+1}+1/2}$,
using $\rho_{\Lambda w}$ in Eq. \ref{chemipotezerotemp}.
 Comparing with $n_\Lambda \Gamma_1\sim  \rho_\Lambda^{3/2}, $
from  Eq. \ref{Gamdos}, for $-1<w_\Lambda< 1$, $\Gamma_2
\ll\Gamma_1$ as $\rho_\Lambda\rightarrow 0$.
 Similarly, this will always occur for low $T_{w\chi }\neq 0$, implying still
$\rho_\Lambda\sim \rho_{\Lambda w}$, but high enough for the thermic
contribution to be dominant so\cite{Kolb}   $\sigma \sim(1/M_P^4)
T^2_{w\chi }$.
 Another type of interaction can be dominant for some time, but it
will eventually be overridden by the $\Gamma_1$ term.
 Lower powers of $\rho_\Lambda$,  e. g.,
a  constant decay rate $n_\Lambda\Gamma_2\sim \rho_\Lambda$, could
make a
 significant cosmological
 contribution, but it  would have to be fine-tuned to give the present
 parameters\cite{besproCosmo}.
Thus, the $\Gamma_2$ term can and will be
 neglected.

Under such circumstances,  we use changes  of the form $\partial
N_\Lambda/\partial V=n_\Lambda$ in Eq. \ref{chemipotenExpand}. We
obtain, using Eqs. \ref{EnergyConserv}, \ref{Gamdos},
\begin{eqnarray}\label  {rholamec}
\dot \rho_\Lambda+3(w_\Lambda+1)H
\rho_\Lambda=3[(w_\Lambda+1)+\chi]H \rho_\Lambda.
\end{eqnarray}
 Energy conservation in Eq. \ref{energyeqCons}  demands that
 energy be transferred, which we assume occurs for only another  dominant  $i$
 component in Eq. \ref{Friedmann},
 \begin{eqnarray}\label {RadMat}
\dot \rho_i+3(w_i+1)H \rho_i=-3[(w_\Lambda+1)+\chi ] H \rho_\Lambda.
\end{eqnarray}
 The set of Eqs.  \ref{Friedmann},
 \ref{rholamec}, \ref{RadMat}
 describes a two-fluid system with $\rho_\Lambda$
 decaying out of
equilibrium as is common in many universe processes$\cite{Kolb}$. No
energy transfer is produced for $w_\Lambda+1+\chi=0$, that is, for
the radiation-like case with  $n\mu_{w\chi}=0$ in Eq.
\ref{chemipoten}. We also find (see Eq. \ref{rholamec}) dark-energy
decay for $\chi< 0.$
 A  decaying cosmological constant was
first conceived by  Bronstein$\cite{Bronstein}$   to explain the
universe's time direction, and recent study starts with Ref.
\cite{Ozer}, with various phenomenological decay laws then
considered$\cite{Reuter}$; quintessence models with a similar energy
interchange have also been studied\cite{Amendola}.
 By substituting $H$ in   Eq. \ref{Friedmann} into   Eq. \ref{rholamec},  we  obtain
\begin{eqnarray}\label {RadMatExpli}
 \rho_i=  -{\rho}_\Lambda
 +\frac{{\dot\rho}_\Lambda^2}{ 24
     \pi  \chi^2  {\rho}_\Lambda^2}.  \end{eqnarray}
   Substituting this  into Eq. \ref{RadMat}, we get
\begin{eqnarray}\label  {MasterEq}
   6\,\chi\,\rho_\Lambda\,  \ddot{\rho}_\Lambda  + \left( d_i - 6\,\chi \right) \,
   {\dot\rho_\Lambda}^2
  -24\,\pi \,[ d_i - 3\,\left( 1 + w_\Lambda \right)
]
 \,{\chi}^2\,
   {\rho_\Lambda}^3 =0,
   \end{eqnarray}
where $d_i=3(w_i+1)$. $t$ as   inverse function   of
$\rho_{\Lambda}$  can be integrated, where initially
$\rho_{\Lambda_i}$ at $t_i$
\begin{eqnarray}\label  {Integrated}
t-t_i=  \int_{ \rho_ {\Lambda } }^{\rho_{\Lambda_i} }  d\rho
\left(\frac{d_i+3 \chi}{ 24{{\chi}}^2 \pi[d_i-3(w_\Lambda+1) ]
{{\rho}}^3+
 3({d_i+3 \chi})  {\chi \,C}{\rho}^{2 - \frac {d_i}{3\,\chi}}    }\right )^{\frac{1}{2}}.
   \end{eqnarray}
$C$  accounts for initial conditions for ${\rho_i}$, and we have
chosen the solution for which $R$ increases and $\rho_\Lambda$
decreases. For some $\chi,$ $w_\Lambda$, $t(\rho_\Lambda)$ can be
given explicitly in terms of hypergeometric and elliptic functions.
Using Eqs. \ref{RadMatExpli}, \ref{Integrated}
  one finds
\begin{eqnarray}\label  {rhocrit}
 \rho_c\approx \frac{24{{\chi}}^2 \pi[d_i-3(w_\Lambda+1) ]
{{\rho_{\Lambda}}} +
 ({d_i+3 \chi}) 3\, {\chi \,C}{\rho_{\Lambda}}^{ - \frac {d_i}{3\,\chi}}}{24
     \pi  \chi^2    (d_i+3 \chi)}.
  \end{eqnarray}
 One derives that for $ -d_i/3<\chi<0
$
\begin{eqnarray}\label  {asymptotic}
{\rm lim}_{\rho_ {\Lambda} \rightarrow 0 } \frac{{\rho_
{\Lambda}}}{\rho_c} =\frac {  d_i+3 \chi }{d_i-3(  w_\Lambda+1)}
\end{eqnarray}  within the wide set of
initial conditions   $C\ll \rho_{\Lambda 0} ^{1 + \frac
{d_i}{3\,\chi}}$,
 so  $\Omega_i$ and $\Omega_\Lambda$ will acquire a fixed asymptotic
 value.

\begin{figure}[h]
 \begin{center}
 \includegraphics[scale=0.9]{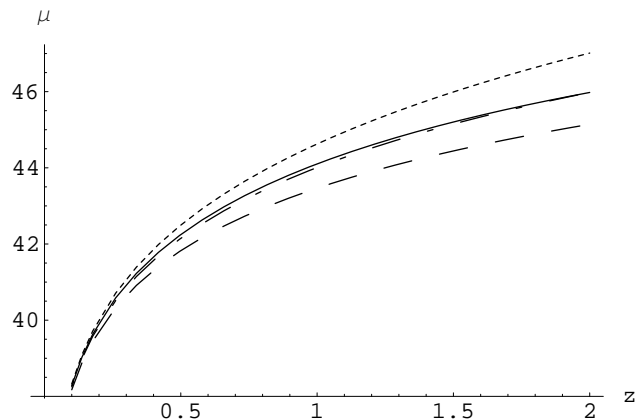}
 \end{center}
 \caption{  Comparison of magnitude $\mu=5 {\rm Log}_{10} (d_L/{\rm Mpc})+25$ of luminosity distance $d_L$,
as a function of redshift $z$, for flat models. For non-asymptotic
models with $w_\Lambda=-1$, and (a) $\Omega_{m0}=0$,
$\Omega_{\Lambda 0}=1$ (dotted), (b) $\Omega_{m 0}=.27$,
$\Omega_{\Lambda 0}=.73$
 (line),
and (c) $\Omega_{m0}=1$, $\Omega_{\Lambda 0}=0$ (dashed); and (d)
for asymptotic model with  $\Omega_{b 0}=.044,$  and $\chi_0=-.48$
(dot-dashed). The reduced Hubble parameter $h=.71$ was used for all
cases.}\label{algun--nombre}
 \end{figure}

Such  an asymptotic behavior fits
 the supernova
data$\cite{Supernova}$  interpreted under Eq. \ref{asymptotic}, with
dark matter and dark energy evolving with a constant ratio.
Considering baryonic matter, dark matter and dark energy (the latter
two evolving as $R^{3\chi_0}$), assuming asymptotic behavior sets in
as early as $z=2$, with the  constant $\chi_0=-.48$, and as shown in
Fig. 1 (and compared with the fitting non-asymptotic model, and
non-fitting $\Omega_{\Lambda 0}=0$, and cosmological-constant
$\Omega_{\Lambda 0}=1 $  cases), one can reproduce the luminosity
distance  $d_L = H^{-1}_ 0 (1 + z) \int ^ z _0 dz^\prime
[\Omega_{b0}(1 + z^\prime)^3+ (1 - \Omega_{b0})(1 +
z^\prime)^{-3\chi_0}]^{-1/2}$ up to the measured redshift $z\sim 2$.
We note that  the fit is independent of $\Omega_{\Lambda 0}$, as
derives from the asymptotic regime. The choice of initial conditions
($C$ in Eq. \ref{Integrated})  sets the timing of the
matter-dominated regime ($w_i=w_m=0$ in Eq. \ref{RadMat})
 before the asymptotic one, to
    match  the conventional cosmology.


\section{Conclusions}

In  summary,
account of  dark energy's quanta allows for a dark-energy decaying
model able to explain its coincidence with dark matter today, within
classical general relativity and thermodynamics. It represents a
departure from the zero-temperature cosmological constant, while it
maintains
 the results of the standard cosmology.
 This supports
 a conservative approach   in   which known physical elements  can provide new
  information$\cite{JaimeLett}$.
Dark energy's coincidence  with the critical density today  is
connected to the universe evolution,
in which events occur by contingency, rather than chance. While
microphysics$\cite{Adler}$ needs to elucidate
 the dark energy's
 equation of state,
the universe  already emerges as
 flat,
 interconnected, evolving deterministically, and  in an inexorable process of
 accelerated expansion and decay.

\section*{Acknowledgement} The author thanks A. de la Macorra and M.
    Giovannini for discussions, and acknowledges support from
    DGAPA-UNAM and HELEN.

\end{document}